\documentstyle[12pt]{article}

\begin{document}
\newcommand {\beqa}{\begin{eqnarray}}

\newcommand {\eeqa}{\end{eqnarray}}
\newcommand{\n}{\nonumber \\}
\newcommand {\beq}{\begin{equation}}
\newcommand {\eeq}{\end{equation}}
\newcommand{\om}{\omega}\newcommand {\s}{\sigma}
\newcommand{\de}{\delta}
\newcommand {\pa}{\partial}
\newcommand {\e}{\epsilon}
\newcommand {\th}{\theta}
\newcommand {\ga}{\gamma}
\newcommand {\Ga}{\Gamma}
\newcommand {\al}{\alpha}
\newcommand {\be}{\beta}
\newcommand {\la}{\lambda}

\begin{titlepage}
\strut\hfill \vspace{.5in}
\begin{center}

\LARGE \bf Principal chiral model scattering and the
alternating
quantum spin chain\\[0.5in]


\large \bf Anastasia Doikou \footnote{e-mail: ad22@york.ac.uk} \\

\normalsize{\it Department of Mathematics, University of York,
Heslington\\ YO10 5DD, York, United Kingdom}\\
\normalsize {and}\\

\large \bf Andrei Babichenko \\

\normalsize{\it Shahal
str. 73/4, Jerusalem, 93721, Israel}\\

\end{center} \vspace{0.5in}

\begin{abstract}
We consider the critical alternating quantum spin chain with
${q_{+}\over 2}$, ${q_{-} \over2}$ spins. Using the Bethe ansatz
technique we find explicit expressions for the $S$-matrix of the
model. We show that in the limit that $q_{\pm} \rightarrow \infty$
our results coincide with the ones obtained for the principal
chiral model level one, for the LL (RR) LR scattering. We also
study the scattering of the bound states of the model and we
recover the results of the XXZ (sine -Gordon) model.
\end{abstract}

\end{titlepage}

\newpage
\section{Introduction}

 In the famous paper \cite{PW} the principal
chiral model (PCM) with $SU(2)$ symmetry with Lagrangian density
\begin{equation}
L_0 = \frac{1}{2\alpha_0} tr
(\partial_{\mu}g^{-1}(x)\partial^{\mu}g(x)) \label{PCM}
\end{equation}
were firstly shown to be equivalent to a Gross-Neveu (GN)
fermionic model
\begin{equation}
\tilde{L_0} = \sum_{a=1}^{N}
i\bar{\psi}_a\gamma_{\mu}\partial^{\mu}\psi_a +
\lambda_0\left(\sum_{a=1}^{N}
\bar{\psi}_a\gamma_{\mu}\tau^{i}\psi_a\right)^2 \label{GN}
\end{equation}
in the limit of infinite number $N$ of fermionic colors
($\tau^i$are Pauli matrices here). Secondly, in order to avoid
ambiguities related to the axial anomaly, the fermionic model was
conjectured to be equivalent to the single fermionic field with
isotopic spin $S$ ($N=2S$). The lattice analogue of the last model
is the spin $S$ quantum spin chain exactly solvable by the Bethe
ansatz method. This scheme was successfully applied to PCM and GN
models with different symmetries \cite{W1}, \cite{ORW},
\cite{PW1}, \cite{R}. Another modification of this scheme can be
used for analysis of the PCM with topological term \cite{PW1}
\begin{equation}
L_1 = {1 \over 2 \al_{0}} tr
(\partial_{\mu}g^{-1}(x)\partial^{\mu}g(x)) +
n\epsilon_{\mu\nu}\int_0^1 dtd^2x~ tr\left(
g^{-1}\dot{g}g^{-1}\partial^{ \mu}g g^{-1}\partial^{\nu}g\right)
\label{WZW}
\end{equation}
where in the second term $g(x,t)$ interpolates between $g(x,0)=1$
and $g(x,1)=g(x)$ and $n$ is an integer. This model was shown to
be equivalent to fermionic GN model in the limit of infinite
number of fermions, with broken chiral symmetry, i.e. different
number of colors for left and right fermions:
\begin{equation}
\tilde{L_1} =
\sum_{a=1}^{q_+} u_a^{\dag}\partial_{+}u_a +
\sum_{a=1}^{q_-} v_a^{\dag}\partial_{-}v_a +
\lambda_0\sum_{a=1}^{q_+} \sum_{b=1}^{q_-} \left(
u_a^{\dag}\lambda^i u_a\right) \left( v_b^{\dag}\lambda^i
v_b\right) \label{GNch}
\end{equation}
where $u_a(v_a)$ are the left(right) Dirac fermions,
$\lambda_0=-\frac{1}{2\alpha_0}$, $\lambda^i$ - the generators of
symmetry group and the difference of fermionic color $a$ numbers
is $q_+ - q_- = n$. Here we are focused on the case where $q_{+}-
q_{-} = 1$, which corresponds to the PCM level one. The
corresponding statistical model for this case is the critical
alternating spin chain with $ {q_{+} \over 2}$, $ {q_{-} \over 2}$
spins (see e.g. \cite{PW1}). We are going to find the explicit
$S$-matrix for the spin chain, using the Bethe ansatz method, and
we will reproduce the proposed massless $S$-matrix for the model
(\ref{WZW}), see \cite{ZAZ}.


\section{Bethe ansatz equations}
\subsection{Ground state and excitations}

We focus here on the alternating spin chain with ${q_{+} \over
2}$, ${q_{-} \over 2}$ spins. In order to introduce a mass scale
to our system we consider the chain with inhomogeneities
\cite{FR}, \cite{BT}, namely
\begin{equation} \omega _{j} = (-)^{j} {1 \over \alpha_{0}}\,,
 \end{equation}
  where $j$ is
$q_{\pm}$ and let us also consider that $q_{+}$ is even (obviously
$q_{-}$ is odd). In other words in the sites of the chain with
${q_{+} \over 2}$ spin $\omega _{q_{+}} = {1 \over \alpha_{0}}$
whereas in the sites with ${q_{+} \over 2}$ spin $\omega _{q_{-}}
= -{1 \over \alpha_{0}}$.
 Following the standard Bethe ansatz
technique \cite{R}, \cite{FR}-\cite{B} for the model (\ref{WZW}),
one gets Bethe equations in the form
\begin{equation}
e_{q_+}(\lambda_\al-{1 \over
\alpha_{0}})^{N_+}e_{q_-}(\lambda_\al+{1 \over \alpha_{0}})^{N_-}=
-\prod_{\beta=1}^{M} e_2(\lambda_\al-\lambda_\be) \label{BE}
\end{equation}
where
\begin{equation}
e_n(\lambda; \nu)=\frac{\sinh \mu(\lambda+{in\over 2})}{\sinh
\mu(\lambda-{in \over 2})} \nonumber
\end{equation}
where $\nu = {\pi \over \mu}$ is the anisotropy parameter. The
spin, energy and momentum of a state are characterized by the set
of quasi particles with rapidities (BA roots) {$\lambda_\al$},
\cite{FT}, \cite{KR}, \cite{FR}, \cite{DMN}
\begin{equation}
E=-{\mu \over 2 \pi }\sum_{j=1}^{M} \sum_{n = q_{+}, q_{-}} { \sin
\mu n \over \sinh \mu (\lambda_{j}-{(-)^{n} \over \alpha_{0}}+
{in\over 2})\sinh \mu (\lambda_{j}- {(-)^{n} \over \alpha_{0}}-
{in \over 2})} \label{energy}
\end{equation}
\begin{equation}
P= {1 \over 2 i} \sum_{j=1}^{M} \sum_{n = q_{+}, q_{-}} \log
{\sinh \mu (\lambda _{j}-{(-)^{n} \over \alpha_{0}}+ {in \over 2})
\over \sinh \mu ( \lambda_{j}-{(-)^{n} \over \alpha_{0}}-{in \over
2})}\label{EPS}
\end{equation}
\begin{equation}
S^z=N_{-}q_- + N_{+}q_+ -M\\. \label{spin}
\end{equation}
We consider here the special case where $N_{+} = N_{-} = N$. We
are interested in the limit $N_\pm,q_\pm,\nu \rightarrow \infty$.
The thermodynamic limit $N_\pm \rightarrow \infty$ of equation
(\ref{BE}) can be studied by standard methods. The string
hypothesis says that solutions of (\ref {BE}) in the thermodynamic
limit are grouped into strings of length $n$ with the same real
part and equidistant imaginary parts
\begin{equation}
\lambda_\al^{(n,j)}=\lambda_\al^n + {i\over 2}(n+1-2j)+i\frac{\pi}{4\mu}
(1-v),~~~j=1,2,...,n \label{STR}
\end{equation}
where $\lambda_\al^n$ is real and $v=\pm 1$ determines the parity
of the string (positive negative parity string). We assume that
$\nu > q_{\pm}$. The ground state consists of two Dirac ``filled
seas'' by strings of length $q_{\pm}$. Therefore the Bethe ansatz
equations for the ground state become,
\begin{equation}
\prod _{j= q_{+}, q_{-}}X_{nj}(\lambda_\al^n-{(-)^{j} \over
\alpha_{0}})^{N}= \prod_{j= q_{+}, q_{-}} \prod_{\beta=1}^{M_{j}}
E_{nj}(\lambda_{\al}^{n}-\lambda_\be^j) \label{K}
\end{equation}
where $n$ can be $q_{\pm}$, and
\begin{eqnarray}
X_{nm}(\lambda)&=& e_{|n-m+1|}(\lambda) e_{|n-m+3|}(\lambda)\ldots
e_{(n+m-3)}(\lambda)e_{(n+m-1)}(\lambda)
\nonumber\\
E_{nm}(\lambda)&=& e_{|n-m|}(\lambda)e_{|n-m+2|}^{2}(\lambda)
\ldots e_{(n+m-2)}^{2}(\lambda)e_{(n+m)}(\lambda).
\end{eqnarray}
For what follows it is necessary to
introduce the following
notations
\begin{equation}
g_{n}(\lambda;\nu) = e_{n}(\lambda \pm
{i \pi\over 2 \mu})={\cosh
\mu(\lambda +{i\over 2}) \over
\cosh \mu(\lambda -{i\over 2})}\\,
\end{equation}
\begin{equation}
G_{nm}(\lambda)=
g_{|n-m|}(\lambda)g_{|n-m+2|}^{2}(\lambda) \ldots
g_{(n+m-2)}^{2}(\lambda)g_{(n+m)}(\lambda)\\,
\end{equation}
\begin{eqnarray}
a_{n}(\lambda; \nu) = {1\over 2 \pi} {d \over d\lambda} i \log
e_{n}(\lambda; \nu),~~~b_{n}(\lambda; \nu) = {1\over 2 \pi} {d
\over d\lambda} i \log g_{n}(\lambda; \nu)
\end{eqnarray}
and the Fourier transforms of $a_{n}$,
$b_{n}$ are given by
\begin{equation}
\hat a_{n}(\omega; \nu) = {\sinh \Bigl((\nu -n){\omega \over 2}
\Bigr ) \over \sinh ({\nu \omega \over 2})} ~~0<n<2\nu,
\end{equation}
\begin{eqnarray}
\hat b_{n}(\omega; \nu) = -{\sinh ({n\omega \over 2}) \over \sinh
({\nu \omega \over 2})} ~~0<n< \nu,\ = -{\sinh \Big ( (n-2
\nu){\omega \over 2} \Bigr ) \over \sinh ({\nu \omega \over 2})}
~~\nu<n<3\nu.
\end{eqnarray}
We also need the following
expressions
\begin{equation}
\Bigl (Z_{nm}(\lambda),\ A_{nm}(\lambda),\ B_{nm}(\lambda)\Bigr )=
{1 \over 2\pi}{d \over d\lambda} i \log \Bigl (X_{nm}(\lambda),\
E_{nm}(\lambda),\ G_{nm}(\lambda)\Bigr )\,,
\end{equation}
where the Fourier
transforms of the last expressions are
\begin{equation}
\hat Z_{nm}(\omega)= { \sinh \Bigl (( \nu - \max(n,m)){\omega
\over 2} \Bigr ) \sinh \Bigl ((\min(n,m)){\omega \over 2} \Bigr )
\over \sinh ({\nu \omega \over 2}) \sinh({\omega \over 2})}\\,
\end{equation}
\begin{equation}
\hat A_{nm}(\omega)={2 \coth ({\omega \over 2} )\sinh \Bigl (( \nu
- \max(n,m)){\omega \over 2}\Bigr) \sinh \Bigl ((\min(n,m)){\omega
\over 2}\Bigr ) \over \sinh ({\nu \omega \over 2})}-\delta_{nm}\\,
\end{equation}
\begin{equation}
\hat B_{nm}(\omega)=-{2\coth ({\omega \over 2})\sinh\Bigl
({n\omega \over 2}\Bigr ) \sinh \Bigl ({m\omega \over 2}\Bigr)
\over \sinh ({\nu \omega \over 2})}. \label{fourier}
\end{equation}
Finally, the energy (\ref{energy}) takes
the form
\begin{equation}
E = - \sum_{i, j=q_{+}, q_{-}}\sum_{\al =1}^{M_{i}}Z_{ij}
(\lambda_{\al}^i-{(-)^{j} \over \alpha_{0}})\\. \label{energyp}
\end{equation}
Now we are going to solve the Bethe ansatz equations. We take the
logarithm and the derivative of the Bethe ansatz equations and we
also use the Maclaurin expansion
\begin{equation}
\sum_{j=1}^{N} f(\lambda_{j}) \sim N
\int_{-\infty}^{\infty}
f(\lambda) \sigma(\lambda)d\lambda
- \sum_{j=1}^{u}f(\lambda_{j})\\,
\label{mac}
\end{equation}
where $\sigma$ is the density of the
state and $u$ is the number
of holes. Then for the
densities of the ground state (no holes) we
obtain the
following integral equations,
\begin{equation}
\sigma_{0}^{n}(\lambda) = \sum_{j= q_{+}, q_{-}}
Z_{nj}(\lambda-{(-)^{j} \over \alpha_{0}}) - \sum_{ j= q_{+},
q_{-}} (A_{nj}*\sigma_{0}^{j})(\lambda) \label{density}
\end{equation}
where $*$ stands
for the
convolution and $n$ can be $q_+$ or $q_-$. We can easily solve
the last equations (\ref{density}) and we find that
\begin{equation}
\sigma_{0}^{n} (\lambda) = s(\lambda-{(-)^{n} \over
\alpha_{0}})={1 \over 2 \cosh\Bigl (\pi(\lambda-{(-)^{n} \over
\alpha_{0}})\Bigr)}\,, \label{energy2}
\end{equation}
where its Fourier transform is
\begin{equation}
\hat s(\omega)= \sum_{i, j=q_{+}, q_{-}} (\hat Z_{ij} \hat
R_{nj})(\omega)= {1\over2 \cosh({\omega \over 2})}\\.
\label{energy0}
\end{equation}
Here $R$ is the inverse of the kernel
$K$ of the system of the
linear equations (\ref{density}),
in particular,
\begin{equation}
\hat K_{nm}(\omega)=(1 +\hat A_{nm}(\omega)) \delta_{nm} +
\hat
A_{nm}(\omega)(1-\delta_{nm})
\end{equation}
\begin{equation}
\hat R_{nm}(\omega)= {1 \over det \hat K} \sum_{j=q_{+}, q_{-}}((1
+\hat A_{jj}(\omega)) \delta_{nm}(1-\delta_{nj}) - \hat
A_{nm}(\omega)(1-\delta_{nm}))\\,
\end{equation}
where the determinant of $K$ is, in terms of trigonometric
functions,
\begin{equation}
det \hat K = {4 \coth^{2} ({\omega \over 2}) \sinh \Bigl ((\nu -
q_{+}){\omega \over 2}\Bigr ) \sinh (q_{-}{\omega \over 2}) \sinh
({\omega \over 2}) \over \sinh ({\nu \omega \over 2})}\\.
\end{equation}
Now we consider the low lying excitations of the model. There can
be holes in both seas of $q_{\pm}$ strings and also strings of
other length and parity (see equation (\ref{STR})). Let us
consider the state with $u_{\pm}$ holes in the $q_{\pm}$ sea. The
contribution of the holes modifies the densities of the states in
the following way (see (\ref{mac})),
\begin{equation}
\sigma^{n}(\lambda) = \epsilon^{n}(\lambda)+{1 \over N} \sum_{i=
q_{+}, q_{-}} \sum_{\al = 1}^{u_{i}} K_{1}^{ni}(\lambda -
\lambda_{\al}^{i})\\, \label{density1}
\end{equation}
where $\epsilon^{n}(\lambda)=s(\lambda-{(-)^{n} \over
\alpha_{0}})$
\begin{equation}
\hat K_{1}^{ni}(\omega) =
\sum_{j=q_{+}, q_{-}}(\hat A_{ij} \hat R_{nj})(\omega)\\,
\end{equation}
In particular, $K$ in (\ref{density1})
have the following explicit
form in terms of trigonometric
functions
\begin{eqnarray}
\hat K_{1}^{q_{+}q_{+}}(\omega)= {\sinh \Bigl ((\tilde \nu
-2){\omega \over 2}\Bigr ) \over 2 \cosh({\omega \over 2}) \sinh
\Bigl ((\tilde \nu -1) {\omega \over 2}\Bigr )},~~~\hat
K_{1}^{q_{+}q_{-}}(\omega) = {1\over 2 \cosh({\omega \over 2})}
\label{K11}
\end{eqnarray}
with $\tilde \nu = \nu - q_{-}$, and
also
\begin{eqnarray}
\hat K_{1}^{q_{-}q_{+}}(\omega)= {1\over 2 \cosh({\omega \over
2})},~~~\hat K_{1}^{q_{-}q_{-}}(\omega) = {\sinh \Bigl ((q_{+}
-2){\omega \over 2}\Bigr ) \over 2 \cosh({\omega \over 2}) \sinh
\Bigl ((q_{+} -1){\omega \over 2} \Bigr )} \label{K12}
\end{eqnarray}
The energy of the state with $u_{\pm}$ holes in the $q_{\pm}$ seas
is given (see (\ref{energyp})) by
\begin{equation}
E= E_{0} + \sum_{j= q_{+}, q_{-}} \sum_{\al=1}^{u_{j}}
\epsilon^{j}(\lambda_{\al}^{j})\\,
\end{equation}
where $E_{0}$ is the energy of the ground state and
$\epsilon^{\pm}(\lambda)$ is the energy of the hole in $q_{\pm}$
sea (we write $\epsilon^{\pm}$ instead of $\epsilon^{q_{\pm}}$ for
simplicity).
 Finally, we
compute the spin of the holes, and we can see that the spin of a
hole in $q_{+}$ sea is from (\ref{spin}) $s^{+}={f \over 2}$,
where $f$ is just a factor $f = 1+ {q_{+} \over \tilde \nu -1}$
which is obviously 1 for $\tilde \nu \rightarrow \infty$.
Therefore, we consider the spin of the hole to be ${1\over 2}$ for
what follows. The spin of a hole in $q_{-}$ sea is see
(\ref{spin}) $s^{-}=0$. Finally, we conclude that the hole in the
$q_{\pm}$ sea is a particle like excitation with energy
$\epsilon^{\pm}$, momentum $p^{\pm}$ ($e^{n}(\lambda) = {1\over
\pi} {d \over d\lambda}p^{n}(\lambda)$)
\begin{equation}\epsilon^{n}(\lambda) = {1 \over 2 \cosh\Bigl (\pi(\lambda-{(-)^{n} \over
\alpha_{0}})\Bigr)}\,, \qquad p^{n}(\lambda)= - {\pi \over 4} +
{1\over 2} \tan^{-1}\Bigl (\sinh\pi (\lambda-{(-)^{n} \over
\alpha_{0}})\Bigr )\,, \label{momentum}
\end{equation}
 and spin $s^{+}={1 \over 2}$, $s^{-} =
0$. We can easily check that in the scaling limit where $\lambda
<< {1\over \alpha_{0}}$ the energy and momentum become from
(\ref{energy2}) and (\ref{momentum}), see also \cite{FR},
\cite{BT},
\begin{equation}
\epsilon^{+}(\lambda) = p^{+}(\lambda) \sim  e^{-{\pi \over
\alpha_{0}}} e^{\pi \lambda}\,, \qquad \epsilon^{-}(\lambda) =
-p^{-}(\lambda) \sim  e^{-{\pi \over \alpha_{0}}}
 e^{-\pi \lambda}\,. \end{equation}
 These are the energy
  and momentum of the ``right'' and ``left'' movers
 respectively  (see e.g. \cite{ZAZ}). The factor
$e^{-{\pi \over \alpha_{0}}}$ provides a mass scale for the
system.

 As we mentioned before there
are also states that consist of strings. There exist two types of
strings as we see from equation (\ref{STR}) (positive and negative
parity). The presence of the string gives rise to some extra terms
in the Bethe equations which give the 1/N contribution to the
density, in particular the Bethe ansatz equations in the presence
of a positive parity $l$-string become,
\begin{equation}
\prod_{j= q_{+}, q_{-}} X_{nj}(\lambda_\al^n-{(-)^{j} \over
\alpha_{0}})^{N}= \prod_{j= q_{+}, q_{-}} \prod_{\beta=1}^{M_{j}}
E_{nj}(\lambda_\al^n-\lambda_\be^j)
E_{nl}(\lambda_\al^n-\lambda_0) \label{K1}
\end{equation}
where $\lambda_0$ is the real center of
the string. The densities
of the state are
\begin{equation}
\sigma^{n}(\lambda) = \sum_{j=q_{+}, q_{-}}Z_{nj}(\lambda-{(-)^{j}
\over \alpha_{0}})- \sum_{j=-,+} (A_{nj} * \sigma^{j}) (\lambda)-
{1 \over N} A_{nl}(\lambda - \lambda_{0})\\,
\end{equation}
we solve
the system of the linear equations and we find that:
\begin{equation}
\sigma^{n}(\lambda) = \epsilon^{n}(\lambda)-{1 \over N}
K_{+}^{nl}(\lambda - \lambda_{0})\\, \label{density+}
\end{equation}
where
\begin{equation}
\hat
K_{+}^{nl}(\omega)= \sum_{j=q_{+}, q_{-}}
(\hat A_{lj}
\hat R_{nj})(\omega)\\.
\label{K+}
\end{equation}
We have to consider two different
cases for the positive parity strings. In particular, we
choose
the following strings, namely, the $l=q_{+}+1$ and
$l=q_{-}-1$.
For each string we obtain different densities,
i.e. for the
$l=q_{+}+1$ string,
\begin{eqnarray}
\hat K_{+}^{q_{+}(q_{+}+1)}(\omega) = -{\sinh \Bigl ( (\tilde \nu
-2){\omega \over 2}\Bigr ) \over \sinh \Bigl ((\tilde \nu
-1){\omega \over 2}\Bigr )},~~~\hat
K_{+}^{q_{-}(q_{+}+1)}(\omega)=0 \label{K21}
\end{eqnarray}
and for $l=q_{-}-1$
\begin{eqnarray}
\hat K_{+}^{q_{+}(q_{-}-1)}(\omega)=0,~~~\hat
K_{+}^{q_{-}(q_{-}-1)}(\omega) = - {\sinh \Bigl((q_{+} - 2){\omega
\over 2}\Bigr) \over \sinh\Big((q_{+}-1) {\omega \over 2})\Bigr )}
\label{K22}.
\end{eqnarray}
The
energy of the
state from (\ref{energy}), (\ref{density1}) and
(\ref{mac}), is given by,
\begin{equation} E = -
\sum_{i, j=q_{+}, q_{-}}\sum_{\al =1}^{M
_{i}}Z_{ij}(\lambda_{\al}^i-{(-)^{j} \over \alpha_{0}})
-\sum_{j=q_{+}, q_{-}}Z_{lj}(\lambda_0-{(-)^{j} \over
\alpha_{0}})\\,
\end{equation}
we find that $E=E_{0}$, therefore we conclude that this is a state
with zero energy contribution. The spin (\ref{spin}) of the
$q_{+}+1$ string is $s=-1$ (again we omit the factor $f$ for the
spin) and the spin of the $q_{-}-1$ string is zero. Finally, we
study the state with an one negative parity string. The Bethe
ansatz equations then become
\begin{equation}
\prod_{j=q_{+}, q_{-}} X_{nj}(\lambda_\al^n-{(-)^{j} \over
\alpha_{0}})^{N}= \prod_{j=q_{+}, q_{-}} \prod_{\beta=1}^{M_{j}}
E_{nj}(\lambda_\al^n-\lambda_\be^j)
G_{n1}(\lambda_\al^n-\lambda_{0}) \label{K2}
\end{equation}
Similarly to the previous case the
densities of the state are
\begin{equation}
\sigma^{n}(\lambda) = \epsilon^{n}(\lambda)-{1 \over N}
K_{-}^{n1}(\lambda - \lambda_{0})\\, \label{density-}
\end{equation}
where
\begin{equation}
\hat K_{-}^{n1}(\omega)= \sum_{j=
q_{+}, q_{-}}(\hat B_{1j} \hat R_{nj})(\omega)\\.
\label{K-}
\end{equation}
Then the ${1\over N}$
contribution of the
densities becomes
\begin{eqnarray}
\hat K_{-}^{q_{+}1}(\omega)={\sinh({\omega \over 2}) \over \sinh
\Bigl( (\tilde \nu -1){\omega \over 2}\Bigr)},~~~\hat
K_{-}^{q_{-}1}(\omega) = 0 \label{K-p}.
\end{eqnarray}
The energy for the string is given by
\begin{equation}
E = - \sum_{i, j=q_{+}, q_{-}} \sum_{\al =1}^{M
_{i}}Z_{ij}(\lambda_{\al}^i-{(-)^{j} \over \alpha_{0}})
-b_{q_{+}}(\lambda_{0}-{1 \over \alpha_{0}}) -
b_{q_{-}}(\lambda_{0}+{1 \over \alpha_{0}})\\, \label{energyb}
\end{equation}
and one can show from (\ref{energyb}), (\ref{density-}) and
(\ref{mac}) that the negative string has also zero energy
contribution. Again the spin of this string is $s=-1$. These are
all the necessary states to describe the scattering processes for
the specific model. Note that the inhomogeneities modify only the
energy and the momentum of the model. However, the ${1\over N}$
contributions to the densities, which essentially determine the
scattering amplitudes, remain unaffected by the presence of the
inhomogeneities. In other words the computation of the $S$-matrix
is not affected by the presence or not of inhomogeneities in the
spin chain. Therefore, the $S$-matrix holds true for any
$\lambda$, and we do not have to consider the scaling limit.


\subsection{Physical $S$ matrix}
Having studied the excitations of the model in the previous
section we are ready to compute the complete $S$-matrix. To do so
we follow the formulation developed by Korepin, and later by
Andrei and Destri \cite{K}. First we have to consider the so
called quantization condition.
\begin{equation}
(e^{2iNp}S-1)|\tilde \la_{1}, \tilde \la_{2} \rangle = 0
\label{qc}
\end{equation}
where $p$ is the momentum of the particle, the hole in our case.
For the case of two holes in $q_{+}$ sea (this is the triplet
state $s =1$), we compare the integrated density (\ref{density1})
with the quantization condition. Having also in mind that,
\begin{equation}
\epsilon^{n}(\lambda) = {1 \over  \pi} {d \over d \lambda} \
p^{n}(\lambda)
\end{equation}
we end up with the following expression for the scattering
amplitude
\begin{equation}
S_{0}(\tilde \lambda)= \exp \Bigl \{2 \pi i \sum_{i=1}^{2}
\int_{-\infty}^{\tilde \lambda_{1}}K_{1}^{q_{+}q_{+}}(\lambda-
\tilde \lambda_{i}) d \lambda \Bigr \}
\end{equation}
where $\tilde \lambda = \tilde \lambda_{1} -\tilde \lambda_{2}$
and $\tilde \lambda_{1}, \tilde \lambda_{2}$ are the rapidities of
the holes. In terms of trigonometric functions (\ref{K11})
\begin{eqnarray}
&&S_{0}(\tilde \lambda)=\exp \Bigl \{-\int_{-\infty}^{\infty}
e^{-i \omega \tilde \lambda} {\sinh \Bigl ((\tilde \nu -2){\omega
\over 2}\Bigr ) \over 2 \cosh({\omega \over 2}) \sinh \Bigl
((\tilde \nu -1) {\omega \over 2}\Bigr )}{d\omega \over
\omega}\Bigr \} = \nonumber\\&& \prod_{j=1}^{\infty}{\Gamma({i
\tilde \lambda \over \tilde \nu -1} +{2j-2 \over \tilde \nu -1}+1)
\Gamma({i \tilde \lambda \over \tilde \nu -1} +{2j \over \tilde
\nu -1}) \over \Gamma({i \tilde \lambda \over \tilde \nu -1}
+{2j-1 \over \tilde \nu -1}+1)\Gamma({i \tilde \lambda \over
\tilde \nu -1} +{2j-1 \over \tilde \nu -1})}{\Gamma({-i\tilde
\lambda \over \tilde \nu -1} +{2j-1 \over \tilde \nu
-1}+1)\Gamma({-i\tilde \lambda \over \tilde \nu -1} +{2j-1 \over
\tilde \nu -1}) \over \Gamma({-i\tilde \lambda \over \tilde \nu
-1} +{2j -2\over \tilde \nu -1}+1) \Gamma({-i\tilde \lambda \over
\tilde \nu -1} +{2j \over \tilde \nu -1})}\,. \label{S0}
\end{eqnarray}
To get the singlet we consider the state with two holes in the
$q_{+}$ sea and the $q_{+}+1$ string. Looking at the spins of the
hole in $q_{+}$ sea and the spin of the $q_{+}+1$ string we
conclude that this is indeed a state with spin zero. Note that the
real center of the string is in the middle of the two holes
($\lambda_{0} = {\tilde \lambda_{1} + \tilde \lambda_{2} \over
2}$). We conclude from (\ref{density1}), (\ref{K11}) and
(\ref{density+}), (\ref{K21}) that
\begin{equation}
S_{+}(\tilde \lambda)= \exp \Bigl \{-2 \pi i\int_{-\infty}^{\tilde
\lambda_{1}}K_{+}^{q_{+}(q_{+}+1)}(\lambda-\lambda_{0}) d\lambda
\Bigr \}S_{0}(\tilde\lambda)
 \end{equation}
 and after some
algebra we get,
\begin{equation} S_{+}(\tilde \lambda)=
{\sinh \Bigl ({\pi \over 2(\tilde \nu -1)}(\tilde \lambda +i)\Bigr
) \over \sinh \Bigl ({\pi \over 2(\tilde \nu -1)}(\tilde \lambda
-i)\Bigr )}S_{0}(\tilde \lambda)\\. \label{S+}
\end{equation}
However, we need another
state of spin
zero see \cite{DN}, and this is the state
with two $q_+$ holes and
one negative parity string with
its real center in the middle of
the holes. For which we
obtain (\ref{density-})-(\ref{K-p})
\begin{equation}
S_{-}(\tilde \lambda)={\cosh \Bigl ({\pi \over 2(\tilde \nu
-1)}(\tilde \lambda +i)\Bigr ) \over \cosh \Bigl({\pi \over
2(\tilde \nu -1)}(\tilde \lambda -i)\Bigr )} S_{0}(\tilde
\lambda)\\. \label{S-}
\end{equation}
The equations (\ref{S0}), (\ref{S+}) and (\ref{S-}) give the XXZ
$S$-matrix (sine -Gordon $S$-matrix, provided that $\beta^{2} = 8
\pi(1 - {1\over \tilde \nu})$ and $\tilde \lambda = {\theta \over
\pi}$) \cite{BT}, \cite{KR}, \cite{DN}, \cite{ZZ}. Moreover, we
have holes in the $q_{-}$ sea, which are particles with zero spin.
The scattering between two such holes (see (\ref{density1}),
(\ref{K12})) is described by the following $S$-matrix
\begin{eqnarray}
S'_{0}(\tilde \lambda)=\exp \Bigl \{-\int_{-\infty}^{\infty} e^{-i
\omega \tilde \lambda} {\sinh\Bigl ((q_{+} -2){\omega \over
2}\Bigr) \over 2 \cosh({\omega \over 2})
\sinh\Bigl((q_{+}-1){\omega \over 2}\Bigr)}{d\omega \over \omega}
\Bigr \}\,,
\end{eqnarray}
which, in the limit
$q_{\pm} \rightarrow \infty$, corresponds
to the XXX
scattering amplitude \cite{FT}, namely,
\begin{equation}
S'_{0}(\tilde \lambda)=\exp \Bigl \{-\int_{-\infty}^{\infty} e^{-i
\omega \tilde \lambda} {e^{-{\omega \over 2}} \over 2
\cosh({\omega \over 2})}{d\omega \over \omega} \Bigr
\}={\Gamma({i\tilde \lambda \over 2 }) \Gamma({1\over 2}-{i\tilde
\lambda \over 2} ) \over \Gamma({-i\tilde \lambda \over
2})\Gamma({1\over 2}+{i\tilde \lambda \over 2})}\\. \label{S'}
\end{equation}
If we add a $q_{+}+1$ string in the above state the only thing
that changes is the spin of the state which becomes -1, the
scattering amplitude remains unaffected. There is one more state
which completes the correspondence to the XXX $S$-matrix: the
state with two holes in the $q_{-}$ sea and a positive parity
$q_{-} -1$ string (spin zero state). For such state the change in
$\sigma^{q_{-}}$ density gives rise to an extra phase, namely,
\begin{equation}
S'_{+}(\tilde \lambda)= {\tilde \lambda + i  \over \tilde \lambda
-i} S'_{0}(\tilde \lambda)\\. \label{S'+}
\end{equation}
The last two equations give the XXX $S$-matrix (see e.g \cite{FT},
\cite{FR}). The equations (\ref{S0}) - (\ref{S'+}) give actually
the correct $S_{XXZ} \times S_{XXX}$ matrix \cite{BT}. In the
limit $\tilde \nu \rightarrow \infty$ the XXZ matrix reduces to
the XXX $S$-matrix, i.e. the matrix given by (\ref{S'}),
(\ref{S'+}). Therefore we end up with the $SU(2)_{L} \times
SU(2)_{R}$ $S$-matrix \cite{PW}, \cite{ZZ}, \cite{DB}. Finally,
let us consider the state that consists of two holes in $q_{+}$
sea and two holes in $q_{-}$ sea (we consider even number of holes
in each sea). The scattering between the two different holes is
described (\ref{density1}), (\ref{K11}),
\begin{equation}
S_{LR}(\tilde \lambda)= \exp \Bigl \{-\int_{-\infty}^{\infty}
e^{-i \omega \tilde \lambda} {1 \over 2 \cosh({\omega \over
2})}{d\omega \over \omega} \Bigr \}\\,
\end{equation}
which can be
written in the following form
\begin{equation}
S_{LR}(\tilde \lambda)= \tanh{\pi \over 2}(\tilde \lambda
-{i\over
2})\\, \label{LR}
\end{equation}
the same result (up to an overall factor) is obtained for the spin
1/2, 1 alternating isotropic spin chain see \cite{DMN}. The last
result (\ref{LR}) agrees with the proposed $S$-matrix for the left
right scattering \cite{ZAZ}. We identified the $SU(2)_{L} \times
SU(2)_{R}$ $S$-matrix and the $S_{LR}$ matrix as well. This is the
proposed $S$-matrix that describes the massless scattering for the
PCM level one.

\section{Bound states}

It is well known for the XXZ model that for specific values of the
anisotropy parameter there exist bound states of ``kink-anti
kink''. The same happens for the alternating spin chain as well.
In particular, for the $S_{XXZ}$ matrix (\ref{S0}), (\ref{S+}),
(\ref{S-}) and for $q_{+}+1 > \nu$, ($\tilde \nu < 2$ which
corresponds to $\beta^{2}< 4 \pi$) there exist poles in the
physical strip, $\lambda_{p} = i- ik(\tilde \nu -1)$, $k=1, 2,
\ldots < {1\over \tilde \nu -1}$ see also e.g. \cite{ZZ} for a
more detailed analysis. We are going to study here the first
breather $k=1$ which corresponds to a negative parity string of
length one. For $\tilde \nu <2$ (obviously here we keep $\tilde
\nu$ finite) the Fourier transform of $B_{1q_{+}}$ changes, namely
\begin{equation}
\hat{B}_{1q_{+}}(\omega)={2\sinh\Bigl((\nu -q_{+}){\omega \over
2}\Bigr) \cosh\Bigl((\nu -1){\omega \over 2}\Bigr) \over
\sinh({\nu \omega \over 2})}\\. \label{fourier2}
\end{equation}
Having that in mind and also the equations (\ref{density-}),
(\ref{K-}) we conclude that the ${1\over N}$ contribution of the
densities for the breather state are,
\begin{eqnarray}
 \hat K_{-}^{q_{+}1}(\omega) = -{\cosh\Bigl((\tilde \nu - 1){\omega \over
2}\Bigr) \over \cosh({\omega \over 2})},~~~\hat
K_{-}^{q_{-}1}(\omega) = {\cosh\Big((\tilde \nu - 2){\omega \over
2}\Bigr) \over \cosh({\omega \over 2})}.
\end{eqnarray}
The negative parity one string for $\tilde \nu <2$ is a state with
positive energy and zero spin and it corresponds to a breather. In
particular the energy of a breather is see (\ref{energyb}),
(\ref{density-})
\begin{equation}
\epsilon_{1}(\lambda) = {1\over2}\Bigl({1\over \cosh \Bigl
(\pi(\lambda -{1\over \alpha_{0}}+{i(\tilde \nu -2)\over 2})\Bigr
)}+{1\over \cosh\Bigl (\pi(\lambda -{1\over \alpha_{0}}-{i(\tilde
\nu -2)\over 2})\Bigr )}\Bigr )\,, \label{energybr}
\end{equation}
where
\begin{equation}
\hat \epsilon_{1}(\omega) = {\cosh\Bigl((\tilde \nu -2){\omega
\over 2}\Bigr) \over \cosh({\omega \over 2})}\\,
\end{equation}
it is indeed positive and it is the same as in the case of the
critical XXZ model in the ``attractive regime'' (see e.g.
\cite{KR}, \cite{DN2}). In the scaling limit $\lambda << {1\over
\alpha_{0}}$ the energy and the momentum of the first breather
become,
\begin{equation}
\epsilon_{1} (\lambda)= p_{1}(\lambda) \sim 2 \sin\Bigl ({\pi
\over 2}(\tilde \nu -1)\Bigr ) e^{-{ \pi \over \alpha_{0}}} e^{
\pi \lambda}\,,
\end{equation}
this is a dispersion relation for massless particle where the
factor $2 \sin({\pi \over 2}(\tilde \nu -1)) e^{-{\pi \over
\alpha_{0}}}$ is the corresponding mass scale. \footnote{In the
XXZ critical model with inhomogeneities $\omega_{j} = (-)^{j}
\Theta$ (see e.g.\cite{BT}, \cite{FR}) one can show that the mass
of the first breather is exactly $2 \sin({\pi \over 2}(\tilde \nu
-1)) 2e^{- \pi \Theta}$, which is also the mass of the sine-Gordon
first breather, provided that $2e^{- \pi \Theta}$ is the ``kink''
mass.}

 We can write the Bethe ansatz equations for a state with
two breathers,
\begin{equation}
\prod_{j =q_{+}, q_{-}} g_{j}(\lambda- {(-)^{j} \over
\alpha_{0}})^{N}= \prod_{j =q_{+},q_{-}} \prod_{\al =1}^{M_{j}}
G_{1j}(\lambda-\lambda_{\al}^j) \prod_{i=1}^{2}e_{2}(\lambda -
\lambda_{i})\\.
\end{equation}
The density will be
\begin{equation}
-\tilde \sigma(\lambda) = b_{q_{+}}(\lambda-{1 \over \alpha_{0}})
+ b_{q_{-}}(\lambda+{1 \over \alpha_{0}}) - \sum_{j= q_{+}, q_{-}}
(B_{1j}* \sigma^{j})(\lambda) - {1 \over N} \sum_{i=1}^{2}
a_{2}(\lambda-\lambda_{i})
\end{equation}
which we can write in the form,
\begin{equation}
\tilde \sigma(\lambda) = \epsilon_{1}(\lambda)+{1 \over N}
\sum_{i=1}^{2} K_{b}(\lambda - \lambda_{i}). \label{tdensity}
\end{equation}
The fourier transform of $B_{1q_{-}}$ is given by equation
(\ref{fourier}), whereas the Fourier transform of $B_{1q_{+}}$ is
given by (\ref{fourier2}). Therefore the Fourier transform of the
$K_{b}$, after some algebra, is
\begin{equation}
\hat K_{b}(\omega)= -{\cosh\Bigl((2 \tilde \nu -3){\omega \over
2}\Bigr) \over \cosh({\omega \over 2})}\\.
\end{equation}
To study the scattering between two breathers we consider the
following quantization condition for a breather
\begin{equation}
(e^{2iNp_{1}}S-1)|\la_{1}, \la_{2} \rangle = 0 \label{qcp}
\end{equation}
where $p_{1}$ is the momentum of the breather, then if we compare
the last equation with (\ref{tdensity})(after we integrate it) we
conclude that
\begin{equation}
S_{b}(\lambda) = \exp \Bigl \{\int_{-\infty}^{\infty} e^{-i \omega
\lambda} {\cosh\Bigl((2 \tilde \nu -3){\omega \over 2}\Bigr) \over
\cosh({ \omega \over 2})}{d \omega \over \omega} \Bigr \}=
{\sinh(\pi \lambda)+ i \sin(\pi(\tilde \nu -1)) \over \sinh(\pi
\lambda ) -i \sin(\pi(\tilde \nu -1))}\\.
\end{equation}
The result agrees with the one obtained in XXZ (sine-Gordon)
model, see e.g. \cite{KR}, \cite{ZZ}, \cite{DN2}. Here, we
restricted ourselves to the case of the ``elementary'' breather
only. The above analysis can be generalized, in a similar way, for
any general n-breather.


\section{Discussion}
We considered the alternating critical quantum spin chain with
${q_{+} \over 2}$, ${q_{-} \over 2}$ spins. We studied the
excitations of the model and their scattering. We showed that in
the limit that $q_{\pm} \rightarrow \infty$ the $S$-matrix we
found coincides with the proposed one \cite{ZAZ} for the PCM level
one. The natural next step is the generalization of the above
method for PCM model at any level. In this case, one expects a
relevance of alternating fused RSOS models by analogy to
\cite{DB}. A Bethe ansatz analysis similar to the one presented
here, was done by Wiegmann \cite{W2} for the reproduction of
massive $S$-matrix for O(3) $\sigma$ model with $\theta = 0$. It
was done on the basis of special parameters limit of the spin S
XXZ chain. In this context the interesting problem is a relation
of the alternating spin chain to the massless $S$-matrix of O(3)
$\sigma$ model with $\theta = \pi$ \cite{ZAZ} and to its
relatives, recently proposed in \cite{F} for other symmetric space
$\sigma$-models with $\theta$ term. It would be also interesting
to study the thermodynamics of generic alternating spin chain (the
case of spin 1/2 spin 1 chain was studied in \cite{AM},
\cite{DMN1}, \cite{DM}) in order to see its relation to the
thermodynamics of the corresponding field theoretical model, which
was studied by methods of TBA. We hope to report on these issues
in a future work.

\section{Acknowledgements}
AB is thankful to P.Wiegmann for initiation of interest to the
problem and for useful discussions at the beginning stage of the
work. One of us AD is supported by an EPSRC postdoctoral
fellowship.



\begin{thebibliography}{99}

\bibitem{PW} A. Polyakov and P. Wiegmann, {\em Phys. Let.}, {\bf B131} (1983) 121.

\bibitem{W1} P. Wiegmann, {\em Phys. Let.} {\bf B142} (1984) 173.

\bibitem{ORW} E. Ogievetsky, N.Yu. Reshetikhin and P. Wiegmann, {\em Nucl. Phys.} {\bf
B280} (1987) 45.

\bibitem{PW1} A. Polyakov and P. Wiegmann, {\em Phys. Let.} {\bf B141} (1984) 223.

\bibitem{R}
N.Yu. Reshetikhin, {\it Nucl. Phys.} {\bf B251} (1985) 565.

\bibitem{ZAZ} A. Zamolodchikov and Al. Zamolodchikov, {\em Nucl. Phys.} {\bf
B379} (1992) 602.

\bibitem{FT} L.D. Faddeev and L.A. Takhtajan, {\em Russ. Math.
Surv.} {\bf 34, 11} (1979); {\em J. Sov. Math.} {\bf24} (1984)
241.

\bibitem {FR} L.D. Faddeev and N.Yu. Reshetikhin, {\em Ann. Phys.}
{\bf 167} (1986) 227.

\bibitem{BT} H. Babujian and A. Tsvelik, {\em Nucl. Phys.} {\bf
B265} (1986) 24.

\bibitem{LT} L.A. Takhtajan, {\em Phys. Lett.} {\bf A87} (1982) 479.

\bibitem{VW} H.J. de Vega and F.
Woyanorovich,
{\em J. Phys.}{\bf A25} (1992) 4499.


\bibitem{KR} A. Kirillov and
N.Yu. Reshetikhin, {\em J. Phys.} {\bf A20} (1987) 1565.

\bibitem{DNW} G. Japaridze, A. Nersessian and P. Wiegmann, {\em
Nucl. Phys.} {\bf B230} (1984) 511.

\bibitem{B}
H. Babujian, {\em Nucl. Phys.} {\bf B215} (1983) 317.

\bibitem{DMN} H.J de Vega, L.
Mezincescu and R.I. Nepomechie, {\em Int. J. Mod Phys.} {\bf B8}
(1994) 3473.

\bibitem{K} V. Korepin, {\em Theor. Math. Phys} {\bf 41} (1979)
953;  N. Andrei and C. Destri, {\em Nucl. Phys.} {\bf B131} (1984)
445.

\bibitem{DN} A. Doikou and R.I.
Nepomechie, {\em J. Phys.} {\bf A31} (1998) L621.

\bibitem{ZZ}
A. Zamolodchikov and Al. Zamolodchikov, {\em Ann. Phys.} {\bf 120}
(1979) 253; A.B. Zamolodchikov, {\em Commun. Math. Phys.} {\bf 55}
(1977) 183.

\bibitem{DB} D. Bernard, {\em Phys.
Lett.} {\bf B} (1992) 78.

\bibitem{DN2} A. Doikou and R.I.
Nepomechie, {\em J. Phys.} {\bf
A32} (1999) 3663

\bibitem{W2} P. Wiegmann, {\em Phys. Let.}
{\bf B152} (1985) 209.

\bibitem{F} P. Fendley, {\em cond-mat/0008372,
hep-th/0101034}.

\bibitem{AM} S.R. Aladim and M.J.
Martins, {\em J. Phys.} {\bf A26} (1993) 7287.

\bibitem{DMN1} H.J. de Vega, L. Mezincescu and R.I. Nepomechie,
{\em
Phys. Rev.} {\bf B49} (1994) 13223.

\bibitem{DM} B.D. Doerfel and S. Meisner, {\em J. Phys.} {\bf
A30}
(1996) 6471.

\bibitem{BR} V.V. Bazhanon, N.Yu.
Reshetikhin, {\em Int. J. Mod.
Phys.} {\bf A4} (1989)
115.

\end{thebibliography}
\end{document}